\documentclass[12pt]{article}

\usepackage{amssymb}

\def\sch{Schwarzschild}

\begin{document}

\title{Upper limit of the energy of the photon and the phase
change of photon to material particles at the \sch\ radius}

\author{C. Radhakrishnan Nair\\Institute of Mathematical
Sciences\\Kariyavattom\\Thiruvananthapuram\\Pin 695\,581\\India\\
E-mail:~nairrcc@rediffmail.com}

\date{}
\maketitle

\begin{abstract}
The concept of \sch\ radius is extended to the photon and the
upper limit imposed on the energy of a photon as a result of the
three characteristics of the photon--the constancy of the
velocity of light, the spin value of $1\hslash$ and the zero rest mass
of the photon---is shown. Further the phase change that occurs
to the photon at the \sch\ radius, from energy to matter as a
result of vacuum fluctuations is indicated.
\end{abstract}

The photon, as an elementary particle, is unique. It is the only
elementary particle of energy whereas there are hundreds of
elementary particles of matter. There are three unique
properties for the photon.
\begin{enumerate}
\item In the free space, the photon always propagates with the
same constant velocity, velocity of light $c$, irrespective of the
energy it possesses.
\item The rest mass of the photon is zero.
\item The photon has a constant spin value of $\hslash$
\end{enumerate}

In this short paper, we shall examine one remarkable consequence
of the above three properties of the photon.

The special theory of relativity postulates an upper
limit~\cite{1} for the velocity of light in free space. We shall
show that the upper limit of the velocity of light puts an upper
limit on the rotational angular velocity and hence on the
total energy possible for a photon.

Relativistic equation for the energy of a particle is given by the
equation
\begin{equation}\label{eqn1}
E^2=p^2c^2+m_0c^2
\end{equation}
where

$E$ = Energy of the particle

$p$ =  Momentum of the particle

$c$ = velocity of light

$m_0$ =  Rest mass of the particle.

For a photon, the rest mass is zero and therefore the energy of
the photon is given by the equation
\begin{equation}\label{eqn2}
E^2=p^2c^2
\end{equation}
It is well known that photon has one degree of freedom of
rotation and two degrees of freedom of translation~\cite{2,3}.
Since the photon has no rest mass, the total energy of the
photon must be stored as rotational kinetic energy and
translational kinetic energy.

The first postulate of relativity~\cite{1} (or the Galilean invariance)
states that no object is aware of its motion when moving with
uniform velocity and light is no exception to this. Therefore,
for the photon itself, moving with the velocity of light, its
kinetic energy due to translatory motion will be zero. But this
does not say that the photon does not exist for itself. The
photon has a kinetic energy of rotation and therefore the photon
can feel its existence. Incidentally, this also shows the
necessity of a nonzero spin value for any particle with zero
rest mass. Therefore, from the point of view of the photon
itself, since the photon has no rest mass, the total energy of
the photon is stored as rotational kinetic energy. We can
associate rotation even to a particle with zero rest mass,
provided the particle has spatial extension.

The energy of the photon is given by the expression
\begin{equation}\label{eqn3}
E=h\nu=\frac{hc}{\lambda}
\end{equation}
where $h$ = the Planck's constant, $\nu$ = the frequency of
the photon, $\lambda$ = the wavelength of the photon.

The wavelength guarantees the extension of the photon. The
expression $E=\frac{hc}{\lambda}$ shows that the greater the
energy of the photon, the shorter its wavelength. This implies
that with the increase in the energy, the photon will be more
and more localized. Can this localization go on ad infinitum?
The idea of infinite amount of energy packed in an infinitely
small volume is quite unphysical and it cannot be true. How is
the property of the space altered by the ever increasing energy
density? Do the constant parameters $\hslash$ and $c$ put an
upper limit for the possible energy of a photon? What are the
consequences of this upper limit on the space time geometry?

The Harward Tower experiment~\cite{4} has clearly demonstrated
that the energy of a photon is increased under gravitational
attraction. Can the energy of the photon be increased
continuously without limit by gravitational attraction? Since the
photon itself cannot feel any increase in the translational
kinetic energy, the increased energy of the photon, when
attracted by a gravitational field, is entirely stored as
rotational kinetic energy. Spin is the quantum mechanical
analogue of classical angular momentum. Since the photon has a
spin of $1~\hslash$, we can attribute an angular momentum to the
photon in the semiclassical analysis.

When $L$, $I$ and $w$ are the angular momentum, moment of
inertia and angular velocity of a particle, we have the
expression
\begin{equation}\label{eqn4}
L=Iw
\end{equation}
When $m$, $r$ and $v$ are the mass, radius and velocity of the
particle respectively, we have
\begin{equation}\label{eqn5}
L=mrv
\end{equation}
Special theory of relativity puts $c$ as the upper limit for the
velocity of any signal. Therefore no particle can rotate with a
tangential linear velocity $v$ greater than the velocity of
light. When we put this limiting value of $c$ to $v$ in
equation~(\ref{eqn5}), we have
$$
L=mrc
$$
When the particle we consider is the photon, L is replaced by
the spin $S$ of the photon.
\begin{equation}\label{eqn6}
S=mrc
\end{equation}
For a photon $S=1~\hslash$
$$
\therefore~1~\hslash=mrc.
$$
Here $m$ is the equivalent mass of the photon due to its spin.
\begin{equation}\label{eqn7}
m=\frac{\hslash}{rc}
\end{equation}
We assume that as in the case of material particles, the concept
of \sch\ radius is valid for the photon also.

The \sch\ radius of a black hole is given by the expression
\begin{eqnarray}
r&=&\frac{2Gm}{c^2}\label{eqn8}\\
\frac{r}{m}&=&\frac{2G}{c^2}\label{eqn9}
\end{eqnarray}
The right hand side of the equation~(\ref{eqn9}) is a constant.
Therefore, for any black hole, the increase in the radius must be
accompanied by an increase in the mass. But when we consider the
photon as a black hole, \sch\ radius is the minimum radius that
a photon can have. In the case of material particles, we can
have black holes of varying masses with the corresponding radii.
This is possible because the right hand side value of
equation~(\ref{eqn9}) can be maintained by proportional increase
in the values of $r$ and $m$. This is impossible for the photon
because for the photon, with the increase in the equivalent
mass, the radius decreases. Therefore, unlike in the case of
material particles, there is a unique value of radius and mass,
for a photon to behave as a black hole.

Substituting the value of the radius $r$ from
equation~(\ref{eqn8}) into equation~(\ref{eqn7}), we get
\begin{eqnarray}
m&=& \sqrt{\frac{\hslash c}{2G}}\label{eqn10}
\end{eqnarray}
This is the expression for Planck mass. Planck mass is usually
defined as the minimum mass corresponding to the \sch\ radius of
a classical particle. Planck mass, in the case of the photon, is
the maximum equivalent mass permitted for the photon.

We have
\begin{eqnarray*}
c &=& 2.998\times10^8~{\rm m~s}^{-1}\\
h &=&6.62\times10^{-34}~{\rm Js}\\
G &=& 6.67\times 10^{-11}~{\rm N~m}^2~{\rm kg}^{-2}
\end{eqnarray*}
Substituting these values we get
$$
m=1.54\times10^{-7}~{\rm kgs}
$$
The maximum energy possible for the photon will be
\begin{eqnarray}
mc^2&=&\sqrt{\frac{\hslash c}{2G}}c^2\label{eqn11}\\
&=&8.61\times10^{22}~{\rm MeV}.\nonumber
\end{eqnarray}
%
The highest energy so far observed~\cite{5new} for a photon is
of the order of $10^{13}$~eV.

A simple dimensional analysis in terms of the Planck's
natural units will be illuminating here. The dimension of energy
is $M L^2T^{-2}$. When we substitute for $M$, $L$ and $T$ in
terms of $c$, $G$ and $h$ we get the expression
\begin{eqnarray}
 E &=& ML^2T^{-2}\nonumber\\
 &=&\sqrt{\frac{h c}{G}}\frac{h G}{c^3}\frac{c^5}{hG}\nonumber\\
 &=&\sqrt{\frac{h c}{c^2}}\label{eqn12}
\end{eqnarray}
Expressions~(\ref{eqn11}) and~(\ref{eqn12}) are identical except
for the geometrical factor of $4\pi$ appearing in
equation~(\ref{eqn11}).

What happens to the photon when the photon reaches its upper
limit of energy, having an equivalent mass equal to the Planck
mass. Energy and matter are two phases of the same physical
entity. Hawking~\cite{5,6} has shown that in the case of
primordial black holes, at the \sch\ radius, spontaneous pair
production takes place. At the upper limit of energy permitted
for the photon, having reached the \sch\ radius, it may
be legitimate to presume that the photon will undergo pair
production and transformed to material particles. It is also an
open question whether the photon at the maximum energy limit can
be disintegrated into many photons with smaller energies and
larger wavelengths in the turbulent sea of vacuum fluctuations.
Of course these are pure speculations, either to be proved or
disproved by future theoretical and experimental studies.

\section*{Acknowledgement}

Discussions with G.~Mohanachandran, T.~E.~Gireesh, K.~Suresh
Babu and M.~Sabir were extremely useful. L.~A.~Ajith has happily
helped me in the preparation of the script.

\end{document}